\documentclass[epj]{svjour}
\usepackage{latexsym}
\usepackage{graphicx}
\usepackage{bm}
\begin{document}
\title{The thermal jamming transition of soft harmonic disks in two dimensions}
\titlerunning{Thermal jamming transition in 2D}
% \subtitle{Do you have a subtitle?\\ If so, write it here}
\author{Moumita Maiti\inst{1} \and Michael Schmiedeberg\inst{2}
}                     % Do not remove
\institute{\inst{1}Institut f\"ur Physikalische Chemie, Westf\"alische Wilhelms-Universit\"at (WWU), Corrensstr. 28/30, 48149 M\"unster, Germany \\
\inst{2}Institut f\"ur Theoretische Physik I, Friedrich-Alexander Universit\"at Erlangen-N\"urnberg (FAU), Staudtstra{\ss}e 7, 91058 Erlangen, Germany}
\date{}%Received: date / Revised version: date}
\abstract{
  By exploring the properties of the energy landscape of a bidisperse system of soft harmonic disks in two dimensions we determine the thermal jamming transition. To be specific, we study whether the ground state of the system where the particle do not overlap can be reached within a reasonable time. Starting with random initial configurations, the energy landscape is probed by energy minimization steps as in case of athermal jamming and in addition steps where an energy barrier can be crossed with a small but non-zero probability. For random initial conditions we find that as a function of packing fraction the thermal jamming transition, i.e. the transition from a state where all overlaps can be removed to an effectively non-ergodic state where one cannot get rid of the overlaps, occurs at a packing fraction of $\phi_G=0.74$, which is smaller than the transition packing fraction of athermal jamming at $\phi_J=0.842$. Furthermore, we show that the thermal jamming transition is in the universality class of directed percolation and therefore is fundamentally different from the athermal jamming transition. 
 % end of PACS codes
} %end of abstract
\maketitle
This is the final author's version of the article.\\ The final version of the article is published as\\ Eur. Phys. J. E {\bf 42} (2019), 38\\ and is available at Springer via\\ http://dx.doi.org/10.1140/epje/i2019-11802-3.

\section{Introduction}
\label{intro}

The dynamics of particulate systems can dramatically slow down if the density is increased or the temperature is decreased (for reviews see, e.g., \cite{BerthierandBiroli,hunter2012}). The properties of this phenomena that is also known as glass transition are the subject of a lot of research and discussions \cite{BerthierandBiroli,hunter2012}. A related phenomena is the transition of an athermal system from a state where all overlaps between particles can be removed by minimizing the energy to a disordered state where overlaps cannot be avoided and that is termed a jammed state \cite{ohern2002,ohern2003}. Interestingly, while the packing fraction of this athermal jamming transition might depend on the starting conditions \cite{Chaudhury2010}, the critical behavior close to the transition is universal \cite{ohern2002,ohern2003,Chaudhury2010}.

Recently, the method employed for athermal jamming in \cite{ohern2002,ohern2003} has been modified in order to explore thermal jamming in three dimensions \cite{maiti2018,corwin2017} for a system with finite-ranged repulsive interactions. While in \cite{corwin2017} the particles during the minimization process are kept sticking together, in \cite{maiti2018} we allowed the rare crossing of energy barriers that is impossible in case of an athermal system. Therefore, while our approach in \cite{maiti2018} neglects thermal fluctuations within the valleys of the energy landscape, rare thermal rearrangement events in principle are possible. Both methods \cite{maiti2018,corwin2017} revealed that in case of spheres with finite-ranged harmonic repulsive interactions there is a spatial percolation transition at a packing fraction of $\phi_G=0.55\pm 0.01$ if random initial configurations are used. Therefore, the corresponding transition takes place at much smaller packing fractions than athermal jamming, which occurs at a packing fraction of $\phi_J=0.639$ in case of similar starting conditions and interactions \cite{ohern2002,ohern2003}. Furthermore, we have shown that due to this spatial percolation the system cannot explore the full energy landscape and therefore effectively is non-ergodic at packing fractions above $\phi_G$ in the limit of small but non-zero probabilities for barrier crossing events \cite{maiti2018}. The thermal jamming transition as transition from a fluid state to a non-ergodic glass state is in the universality class of directed percolation \cite{maiti2018} and is similar to a modified random organization transition \cite{milz2013}. The predictions of our energy landscape exploration method have been shown to be in agreement with simulation results \cite{maiti2018b}.

Here we use the method introduced in \cite{maiti2018} in order to study the thermal jamming transition in two-dimensional bidisperse soft disk systems with finite-ranged harmonic repulsive interactions. We find that the transition packing fraction of thermal jamming $\phi_G\approx 0.74$ is much smaller than the one of athermal jamming that in two dimensions occurs at $\phi_J=0.842$ in case of the same starting conditions and interaction potentials \cite{ohern2002}. Furthermore, as in three dimensions the thermal jamming transition in two dimensions also is in the universality class of directed percolation.

At a first glance the dynamics of glassy systems in two dimensions look fundamentally different from the dynamics of comparable systems in three dimensions \cite{flenner2015}. However, it has been revealed that the differences are due to Mermin-Wagner-like fluctuations \cite{mermin1966}, i.e., long-ranged fluctuations that in two dimensions occur in addition to the glassy dynamics \cite{shiba2016,vivek2017,keim2017}. Since we neglect the fluctuation within energy valleys in this article, we cannot observe any long-ranged density fluctuations and therefore our results are directly related to the pure glassy behavior.

Our article is structured as follows, In sec. \ref{sec:methods} we describe the model system and the employed method in detail. The results are presented in sec. \ref{sec:Results} where we first determine the state diagram and then study the critical behavior. Finally, we conclude in sec. \ref{sec:conc}.

\section{Model system and methods}\label{sec:methods}

\subsection{Bidisperse harmonic soft disks}

We consider a bidisperse system of soft disks. Motivated by the mixture that is often employed, e.g., in \cite{ohern2003}, half of the disks have the diameter $\sigma$ and the other half the diameter $1.4\sigma$. Crystallization is suppressed due to this bidispersity. The discs do not interact if they do not overlap. Overlapping particles repel each other according to the harmonic pair potential $V_{ij} = \epsilon(1-\frac{r_{ij}}{\sigma_{ij}})^2$, where $\epsilon$ sets the energy scale, $r_{ij}$ denotes the distance between the two particle and $\sigma_{ij}=(\sigma_i+\sigma_j)/2$ their average diameter. We employ systems with periodic boundary conditions and system sizes ranging from $10^5$ to $6 \times 10^5$ particles.

\begin{figure}
	\centering
	\includegraphics[width=0.9\linewidth]{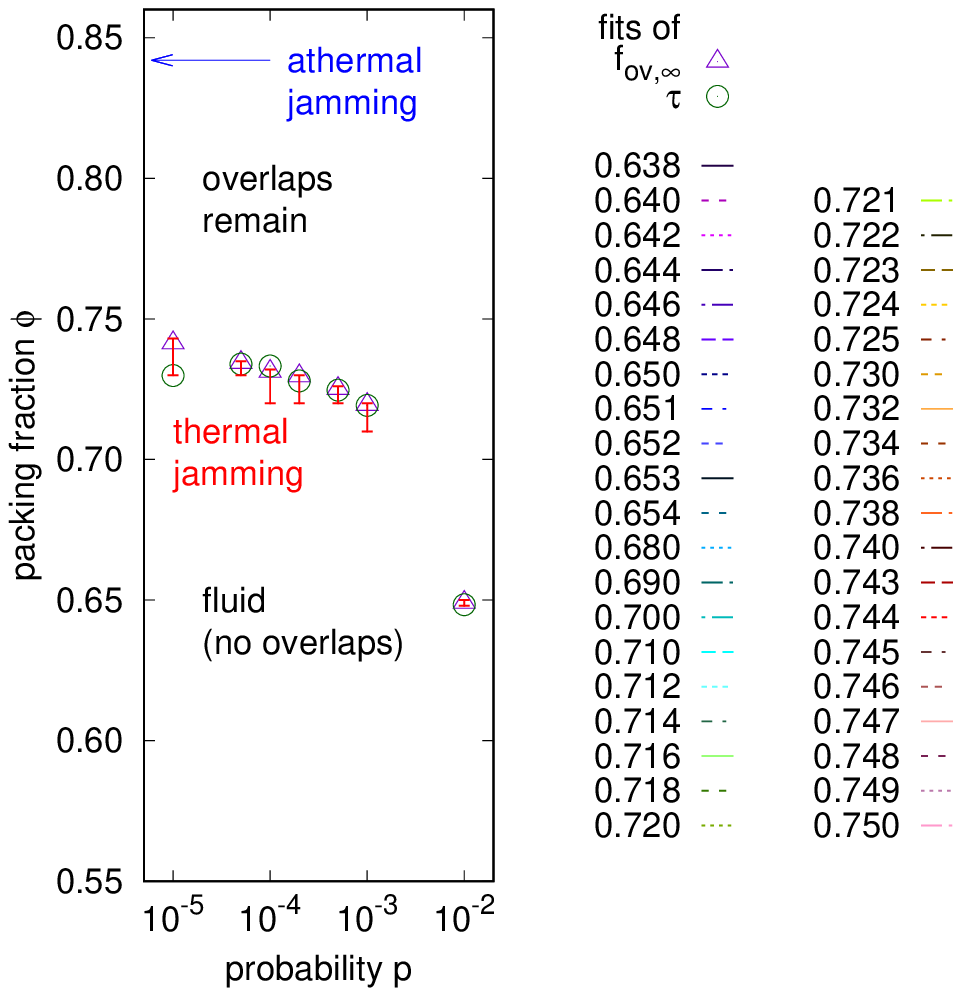}
	\caption{\label{fig:fig1}State diagram as a function of packing fraction $\phi$ and probability $p$ for steps where energy barriers might be crossed. The thermal jamming transition between a fluid state with no remaining overlaps at small packing fractions and a state where overlaps cannot be removed (corresponding to a glass state for small $p$) at large densities is shown by red error bars that indicate the largest observed packing fraction of a fluid and the lowest observed packing fraction of a jammed state not affected by finite size effects in our simulations. The blue arrow indicates the packing fraction $\phi_J=0.842$ \cite{ohern2002} of the athermal jamming transition. The magenta triangles indicate the transition packing fraction obtained by fitting a critical power laws to the fraction of overlaps $f_{ov,\infty}$ at long times. The green circles are obtained from power law fits to the relaxation times $\tau$. The analysis of the critical behavior is explained in detail in sec. \ref{sec:critical}.}
\end{figure}
\begin{figure*}
	\centering
	\includegraphics[width=\linewidth]{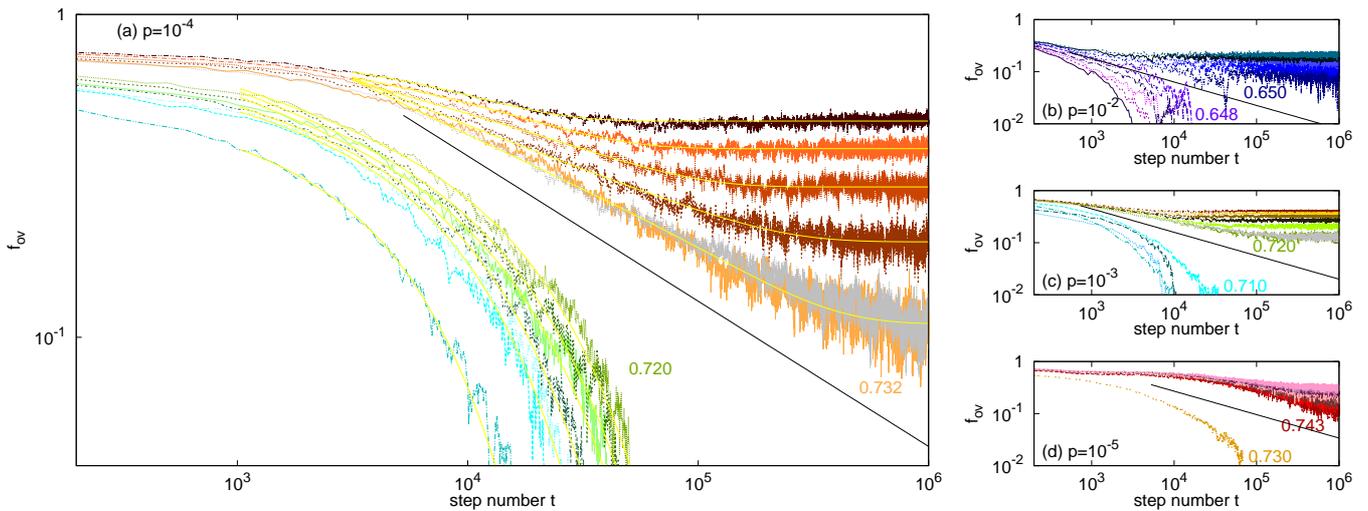}
	\caption{\label{fig:fig2n}Fraction of overlapping particles $f_{ov}$ as a function of steps $t$ of our protocol for various packing fraction. The probability $p$ for possible barrier crossing steps is (a) $10^{-4}$, (b) $10^{-2}$, (c) $10^{-3}$, and (d) $10^{-5}$. The labels in the panel give the packing fractions right below and above the transition. The colors are chosen as in fig.~\ref{fig:fig1}. Each curve corresponds to one quench. In (b) systems with $10^5$, in (a,c) with $N = 5 \times 10^5$, and in (d) with $6 \times 10^5$ particles are considered. We have checked that the behavior of the curves does not change if going to larger or slightly smaller systems. This is demonstrated by the grey curves above the transition in (a,c) that show the results for $6 \times 10^5$ particles. The straight black lines denote the power law $t^{-0.45}$ that is expected at a directed percolation transition \cite{hinrichsen2000}. The yellow curves in (a) are fits according to eq. (\ref{eq:fovt}).}
\end{figure*}

\subsection{Exploration of the energy landscape}

We use the same approach that we have employed for monodisperse systems in three dimensions \cite{maiti2018}. Starting with an random starting configuration we usually employ energy minimization steps that are also used in athermal systems \cite{ohern2002,ohern2003}. However, in each step each particles with an overlap can be selected with a small given probability $p$. A selected particle does not perform a minimization step but is displaced in a randomly chosen direction until it crosses the nearest energy minimum or maximum in that direction. Therefore, such a displacement can lead to the crossing of an energy barrier. The minimization steps or random steps are repeated until all overlaps have been removed or until the fraction of overlapping particles reaches a plateau value and does not further decrease. As for athermal jamming systems that can reach a configuration without overlaps are called unjammed while systems with remaining overlaps are termed jammed.

As discussed in \cite{maiti2018} in the limit of small but non-zero $p$ the observed transition corresponds to a weak ergodicity-breaking transition because in the jammed state the ground state no longer is accessible within a reasonable time. Such an effective ergodicity breaking transition usually is referred to as dynamical glass transition (see. e.g., \cite{BerthierandBiroli,hunter2012}). We want to point out that we consider a dynamical glass transition and not any ideal structural glass transition or the Kauzmann temperature \cite{Kauzmann}. Note that for larger $p$ there are significant rearrangements due to the randomly displaced particles that in principle correspond to ageing but for even large $p$ lead to a thermal fluidization even of the state with remaining overlaps. Therefore, in order to make predictions about the glass transition as a function of the packing fraction, we usually consider the limit of small $p$. For three dimensional systems we have demonstrated that in case of small $p$ this barrier crossing probability can be related to a real temperature $T$ and predictions concerning the temperature-dependence of the glass transition packing fraction are obtained that are in agreement with simulation results \cite{maiti2018b}.

The minimization steps in our protocol are done by using the conjugate gradient algorithm implemented in LAMMPS \cite{lammps}. The minimization is stopped when the energy per particle is equal to $10^{-16}\epsilon$ or smaller and otherwise two particles are considered to overlap (instead of just being in contact) if $\sigma_{ij} - r_{ij}> 10^{-7} \sigma_{ij}$. Note that for three dimensional systems we have tested various modifications of the protocol, e.g., employing steepest descent minimization or using other ways of barrier crossing. However, all modifications led to the same thermal jamming transition \cite{maiti2018}. Furthermore, we have shown that the transition packing fraction can be larger if other starting conditions are employed. Note that this behavior is the well-known history-dependence of the glass transition \cite{BerthierandBiroli}. However the critical behavior does not depend on the initial conditions \cite{maiti2018}.

\section{Results}\label{sec:Results}

For a probability $p=0$, i.e., without barrier crossings the athermal jamming transition is obtained that in case of of random initial conditions occurs at a packing fraction $\phi_J=0.842$ \cite{ohern2002}. In the following we study the thermal jamming transition that occurs for small but non-zero $p$.

\subsection{Thermal jamming transition}

In fig.~\ref{fig:fig1} the state diagram for the thermal jamming transition is shown. The transition between states where all overlaps can be removed at small packing fractions and states with remaining overlaps at large packing fractions is determine by different methods as described in the figure caption. For all methods of analysis, a small transition packing fraction is observed for large probabilities $p$ for barrier crossings and the transition packing fraction increases for decreasing $p$. For small $p$ it stays close to a value of $\phi_G=0.74\pm 0.01$ which is far below the transition packing fraction of athermal jamming that $\phi_J=0.842$ \cite{ohern2002} (indicated by a blue arrow).

In figs.~\ref{fig:fig2n} the relaxation curves for the fraction of overlapping particles $f_{ov}$ as a function of the number of steps $t$ are shown for various packing fractions and probabilities. The employed system sizes are given in the figure caption. Note that we have checked that similar curves are obtained for larger or slightly smaller system sizes. If one wanted to study relaxation curves even closer to the transition, much larger systems would be necessary.

The behavior of the thermal jamming transition is similar to the one that we have found for the three-dimensional systems in our previous work \cite{maiti2018}. For $p$ approaching zero $\phi_{G}$ is the packing fraction where the ergodicity is broken because above this transition the ground state effectively no longer is accessible. Usually this transition is also referred to as dynamical glass transition.
Our result $\phi_G$ is lower than the glass transition density obtained from simulation, where, e.g., a packing fraction of $0.8$ has been reported in \cite{donev2006} for a bidisperse mixture. However, it is larger than the one obtained from mode-coupling theroy, where $0.697$ has been determined in \cite{bayer2007} for a monodisperse packing. Note that we expect that for other starting configurations our approach probably leads to another, usually larger transition packing fraction similar to the behavior that we have observed in three dimensions \cite{maiti2018}.

\subsection{Critical behavior}
\label{sec:critical}

\begin{figure}
	\centering
	\includegraphics[width=\linewidth]{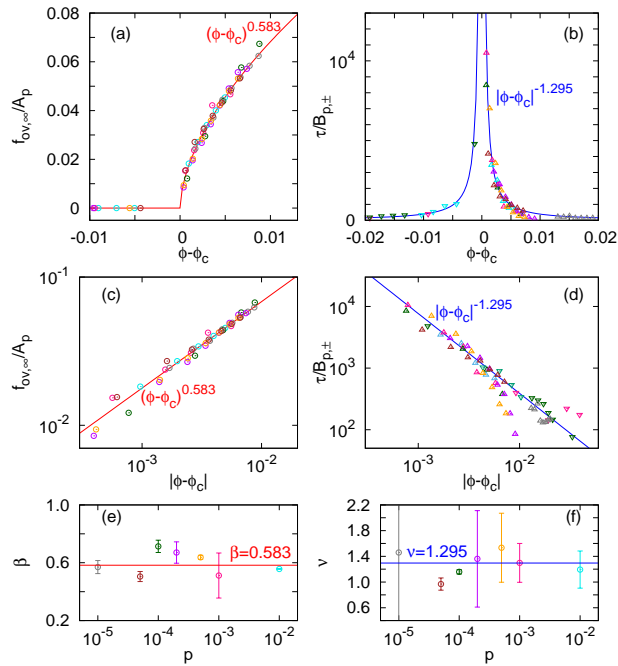}
	\caption{\label{fig:fig2}Analysis of the critical behavior close to the thermal jamming transition. (a,c) Fraction $f_{ov,\infty}$ of particles with overlaps at long times as a function of the difference of the packing fraction $\phi$ from the transition packing fraction $\phi_c$ in (a) linear-linear and (c) log-log-representation. (b,d) Relaxation times $\tau$ as function of $\phi-\phi_c$ in the same representations. Triangles pointing upwards or downwards are used to indicate data above or below the transition, respectively. (e,f) Exponents found by the fitting procedure. Colors in all panels indicate the probability $p$ as can be read of for the points in (e,f). The red and blue lines indicate the power laws or exponents expected for a directed percolation transition \cite{hinrichsen2000}. Note that for (a-d) $f_{ov,\infty}$ and $\tau$ are determined by fits to the relaxation curves as explained in the text. The transition packing fraction $\phi_c$ as well as the rescaling factors $1/A_p$ and $1/B_{p,\pm}$ are obtained by fitting power laws to the data in (a,b). The obtained values of $\phi_c$ are also shown in fig. \ref{fig:fig1}. }	
\end{figure}

In order to analyze the critical behavior close to the transition, we employ the same analysis as in \cite{reichhardt2009,milz2013}. To be specific, we fit the function
\begin{equation}\label{eq:fovt}
f_{ov}(t)=C\frac{e^{-t/\tau}}{t^{\gamma}}+f_{ov,\infty}
\end{equation}
with fit parameter $C$, $\tau$, $\gamma$, and $f_{ov,\infty}$ to the relaxation curves that are exemplary shown in figs.~\ref{fig:fig2n}, i.e., to the fraction of overlapping particles $f_{ov}$ as a function of the numbers of steps $t$. The fitting constant $\tau$ indicates the time until the system reaches a steady state, $f_{ov,\infty}$ is the plateau value of the relaxation curve giving the fraction of remaining overlapping particles in jammed systems. In case the system is not jammed, $f_{ov,\infty}=0$. Close to the transition the relaxation curves can be described by a power law with exponent $\gamma$. 

For $f_{ov,\infty}$ and $\tau$ as functions of the packing fractions, we now fit power laws
\begin{equation}\label{eq:fovu}
  f_{ov,\infty} = A_p (\phi - \phi_{c})^{\beta}
\end{equation}
and 
 \begin{equation}
  \tau = B_{p,\pm} \left|\phi - \phi_{c}\right|^{-\nu}
\end{equation}
for each value of the probability $p$, where $A_p$ and $B_{p,\pm}$ are factors that can depend on $p$ and in case of $B_{p,\pm}$ also on whether one is below or above the transition. The data and a power law curve with the exponents $\beta=0.583$ and $\nu=1.295$ as expected for a directed percolation transition in 2+1 dimensions \cite{hinrichsen2000} are shown in figs.~\ref{fig:fig2}(a-d). The transition packing fractions $\phi_{c}$ determined by fitting are shown in fig.~\ref{fig:fig1}. Note that since we obtain $\phi_{c}$ as a fitting parameter, error bars might be quite large. However, we believe that this is less prejudiced analysis than if we just assumed some values for  $\phi_{c}$. The exponents extracted by fits are displayed in figs.~\ref{fig:fig2}(e,f), where the horizontal lines indicate the mentioned values expected for a directed percolation transition. The error bars indicate the statistical errors obtained from the fitting procedure. However, the true error can be different because small variations of the fitting procedure like excluding the data points that have the largest distance to the transition might lead to similar or even slightly larger changes of the fit value.

Overall, our results are in good agreement with the critical behavior expected for a directed percolation transition and that has been observed for the corresponding three-dimensional system (where the directed percolation transition is in 3+1 dimensions) \cite{maiti2018}. Concerning the two dimensional system considered here, we find that the exponent $\beta$ for $10^{-4}\leq p<10^{-3}$ seem to be slightly larger than the literature value. Though the deviations are outside the statistical error, they are still within the variations that might arrise from different ways of fitting. Note that it is known that for $p=1$ a random organization-like transition is obtained that is in the universality class of directed percolation \cite{milz2013}. Here we do not see any systematic changes of $\beta$ as $p$ is decreased.

Concerning the analysis of the critical behavior related to the relaxation times, we first want to point out that for relaxation curves below the transition, we only have enough data that is sufficiently close to the transition in order to fit a power law for $p=10^{-2}$, $p=10^{-3}$, and $p=10^{-4}$. Above the transition, we have sufficient data for all cases. However, for all data we observe a large scattering. While for small $\left|\phi - \phi_{c}\right|$ the data is in good agreement to the expected power law behavior, there are significant deviations for larger $\left|\phi - \phi_{c}\right|$ as can be seen in the right part of fig.~\ref{fig:fig2}(d). These deviations probably are due to problems that occur when fitting the function of eq.~(\ref{eq:fovt}) to relaxation curves that are never close to the power law-like behavior that is assumed by the $t^{\gamma}$-term in eq.~(\ref{eq:fovt}). The reason that the power laws does not show up (especially for many cases below the transition) is that the relaxation times might be to small, i.e., that we are still not close enough to the transition. Getting closer to the transition would require much larger system sizes that we cannot simulate at the moment within a reasonable time. Note that for the three-dimensional system we have observed the occurrence of an additional power law decay with exponent $-1.5$ that seems to be unrelated to the transition but made it impossible to analyze relaxation times below the transition for all cases \cite{maiti2018}. Here, in the two-dimensional case, we do not find any any additional power law decays and therefore we can analyze the relaxation times below the transition for the previously mentioned cases. The scattering of the data results in large error bars in fig.~\ref{fig:fig2}(f) at least in the cases where a larger number of data points further away from the transition is still part of the analysis.
Nevertheless, though we cannot determine $\nu$ with an accuracy that would be sufficient to rule out other universality classes if we only considered $\nu$ in our analysis, our results show that the critical behavior concerning the relaxation times is in agreement with the critical behavior of a directed percolation transition.

Finally, we look on the power laws with exponents $\gamma$ that are expected close to the transition. Averaging the values of $\gamma$ as determined by fits according to eq.~(\ref{eq:fovt}) over all results for curves with $|\phi-\phi_c|<0.001$, where $\phi_{c}$ is taken from the fit in eq.~(\ref{eq:fovu}), we find $\gamma = 0.44\pm 0.02$, which is close to the literature value $\gamma=\nu/\beta=0.45$ \cite{hinrichsen2000}.

In summary, all exponents $\beta$, $\nu$, and $\gamma$ agree to the critical behavior of a directed percolation transition and there is now systematic change when the probability is changed.

\section{Conclusions}\label{sec:conc}

We have employed our previously introduced approach of exploring the energy landscape in order to study the thermal jamming transition. In this approach we usually minimize the energy but with a small, non-zero probability introduce a step where energy barriers can be crossed. If all overlaps between particles can be removed by this protocol, the system is called unjammed. If the system is stuck in a state with remaining overlaps, this state is termed jammed. In the latter case our protocol was not able to access the ground state of the system within our simulation time and as a consequence all simulation methods that consist of - usually less efficient - energy minimization and thermal fluctuations cannot reach the ground state as well. Therefore, the system effectively is non-ergodic and the thermal jamming transition in the limit of small barrier crossing probabilities corresponds to the dynamical glass transition (cf. discussion in \cite{maiti2018}).

As in our previous work \cite{maiti2018} where we considered a three-dimensional system, the critical behavior of the thermal jamming transition corresponds to the one known for directed percolation transitions.
Furthermore, the thermal jamming transition occurs at a packing fraction that is much smaller than the one of the athermal transition. However, we expect that for different initial conditions the differences can be smaller as we have demonstrated in three dimensions \cite{maiti2018}.

Note that it is well known that the glass transition can occur at packing fractions below the transition packing fraction of athermal jamming (see, e.g., \cite{ikeda,Wang}). The properties of glasses that occur at packing fractions above the glass transition but below athermal jamming have been analyzed in \cite{Wang}. For example, in the hard sphere limit just above the glass transition the glasses can only carry longitudinal but no transverse phonons \cite{Wang}.

An extension of our approach to systems with more complex pair interactions probably would be interesting, e.g., in order to study the temperature-dependence of soft disks \cite{BerthierandWitten,BerthierandWitten1,haxton2011,schmiedeberg2011,maiti2018b} or maybe even the reentrant glass transitions that occur at very large packing fractions \cite{berthier10,schmiedeberg13,miyazaki16}. Recently, a minimization protocol has been used in order to study the jamming transition of attractive systems revealing a second order transition except for weakly attractive systems where a first order transition has been reported that might only occur due to finite size effects \cite{Koeze18}. It would be interesting, to try to study complex gel networks in a similar way \cite{gel,kohl}. Finally, exploring the energy landscape can also be used to study systems that are driven out of equilibrium, e.g., by shearing the system \cite{heuer1,heuer2} or by using self-propelling particles \cite{maiti2018aktiv}.

\begin{acknowledgement}
The project was supported by the Deutsche Forschungsgemeinschaft (Grant No. Schm 2657/3-1). We thank Sebastian Ru\ss\ for helpful discussions and gratefully acknowledge the computer resources and support provided by the Erlangen Regional Computing Center (RRZE).
\end{acknowledgement}

\section*{Authors contributions}
M.M. carried out the simulations and M.S. designed the research and the model system. Both authors analyzed the results and wrote the article.

\end{document}